\documentclass[twoside,10pt]{article}
\usepackage [top=25truemm,bottom=25truemm,left=25truemm,right=25truemm]{geometry}
\usepackage[utf8]{inputenc}
\usepackage[T1]{fontenc}
\usepackage{multirow} 
\usepackage{siunitx} 
\usepackage{graphicx} % Required for inserting images
\usepackage{amsmath}
\usepackage{mathtools}
\usepackage[sort&compress,numbers]{natbib}

\title{Conflict-free joint decision by lag and zero-lag synchronization in laser network}
\author{Hisako Ito${}^*$, Takatomo Mihana, Ryoichi Horisaki, Makoto Naruse}
\date{}
\begin{document}
\columnseprule=0.2mm
\maketitle
\vspace{-2.1\baselineskip}
\begin{center}
{\small Department of Information Physics and Computing, Graduate School of Information Science and Technology,\\
The University of Tokyo, 7-3-1 Hongo, Bunkyo, Tokyo 113-8656, Japan.\\
$^*$Corresponding author. Email: \texttt{ito-hisako780@g.ecc.u-tokyo.ac.jp}
}
\end{center}

\begin{center}\textbf{Abstract}\end{center}\vspace{-0.5\baselineskip}
With the end of Moore's Law and the increasing demand for computing, photonic accelerators are garnering considerable attention.  
This is due to the physical characteristics of light, such as high bandwidth and multiplicity, and the various synchronization phenomena that emerge in the realm of laser physics. These factors come into play as computer performance approaches its limits.
In this study, we explore the application of a laser network, acting as a photonic accelerator, to the competitive multi-armed bandit problem. In this context, conflict avoidance is key to maximizing environmental rewards.
We experimentally demonstrate cooperative decision-making using zero-lag and lag synchronization within a network of four semiconductor lasers. 
Lag synchronization of chaos realizes effective decision-making and zero-delay synchronization is responsible for the realization of the collision avoidance function.
We experimentally verified a low collision rate and high reward in a fundamental 2-player, 2-slot scenario, and showed the scalability of this system.
This system architecture opens up new possibilities for intelligent functionalities in laser dynamics.

\section{Introduction}
At the dawn of the 21st century, Moore's Law\cite{Moore1998}, which predicted consistent improvements in microprocessor performance, was nearing its limit, prompting the exploration of new computing frameworks\cite{Mehonic2022,waldrop2016chips}. 
In particular, the demand for computational power has been increasing dramatically with the use of graphics processing units (GPUs).
As such, GPUs can be viewed as accelerators in computing.
Photonic accelerators\cite{Kitayama2019} attempt to solve specific problems and take advantage of the high speed and multiplicity offered by using light.
In recent years, there have been many studies that apply optics to computational tasks found in photonic accelerators, such as matrix multiplication\cite{zhou2022photonic,feldmann2021parallel}, reservoir computing\cite{Larger2012, Brunner2013}, neural networks\cite{Wagner1987, Marinis2019, Sunada2021, Ohno2022}, neuromorphic computing\cite{shastri2021photonics}, reinforcement learning\cite{flamini2020photonic,bueno2018reinforcement}, coherent Ising machines\cite{Inagaki2016}, and decision-making\cite{Naruse2015, Naruse2017, Homma2019, Mihana2019, Morijiri2023, Iwami2022}.

Decision-making in the realm of the multi-armed bandit (MAB) problem is a complex task. This problem, first described by Robbins in 1952\cite{Robbins1952}, involves choosing from multiple slot machines, each with an unknown probability of winning, with the aim of maximizing profit\cite{Sutton1998}. In this scenario, excessive exploration could result in missed opportunities for winning, while an over-reliance on prior results might lead to the neglect of a potentially profitable machine. As such, successful decision-making requires a delicate balance between exploration and exploitation\cite{auer2002finite,daw2006cortical}. 
Several decision-making strategies for the MAB problem have been demonstrated using photonic principles, leveraging technologies such as single photons\cite{Naruse2015}, chaotic laser\cite{Naruse2017, Naruse2018}, and ring lasers\cite{homma2019chip}. 
The competitive multi-armed bandit (CMAB) problem takes the MAB problem a step further by considering scenarios involving multiple players\cite{Lai2011}. 
This problem encompasses a wide range of issues, including resource competition scenarios like network connection problems\cite{Akkarajitsakul2011}. 
As automation and mechanization continue to evolve and become more widespread, the significance of the CMAB problem grows accordingly. 
In CMAB scenarios, the rewards are divided due to conflicting choices, necessitating conflict-free, cooperative decision-making to optimize both individual and collective profits. This cooperative decision-making has been demonstrated using entangled photons\cite {Chauvet2019, Chauvet2020, Amakasu2021}, instead of single photons. 
In this implementation, the function of photodetectors, where photons are observed, corresponds to slot machine selection.
However, making decisions with constant sampling is challenging due to the randomness of photon emission, and executing these decisions experimentally requires a stable environment. Hence, there is a growing need for a photonic accelerator that offers more flexible scalability and better experimental feasibility for cooperative decision-making.

Given the need for experimental feasibility, decision-making strategies utilizing chaotic lasers have been implemented using learning thresholds and temporal waveforms produced by semiconductor lasers with feedback\cite{Naruse2017}. Chaotic laser systems, compared to single-photon systems, are more conducive to large-scale experiments with fiber optics, such as physical random number generators\cite{Uchida2008, Sakuraba2015}.
Notably, chaotic lasers not only offer fast and complex temporal waveforms but also exhibit synchronization phenomena. Various applications leveraging this synchronization have been reported, including secure communication\cite{Argyris2005}, secure key distribution\cite{Koizumi2013}, and reservoir computing\cite{Larger2012}. 
We have numerically and experimentally implemented a decision-making system using lag synchronization of chaos\cite{Heil2001, Kanno2017, Mihana2019}. This approach expands the possibilities for practical applications of chaotic lasers in the realm of decision-making.
 
The key to this decision-making system is low-frequency fluctuation (LFF) dynamics\cite{Sano1994}.
The temporal waveform in LFF dynamics is characterized by rapid chaotic fluctuations on the order of \SI{}{\giga\hertz} and periodic fluctuation frequency on the order of \SI{}{\mega\hertz}, which consists of a sudden dropout and a gradual recovery phase\cite{ohtsubo2013}.
Thus, LFF embodies dynamics that possess both chaotic and periodic properties. 
The lag synchronization of chaos refers to the phenomenon wherein one laser synchronizes with another laser, with a time delay equivalent to the propagation delay time, denoted by $\tau$.
In this lag synchronization, a laser with an advanced oscillation is called the ``leader,'' while a laser that synchronizes with another laser is called the ``laggard.''
Interestingly, this leader--laggard relationship in LFF dynamics spontaneously switches with a delay time $\tau$ \cite{Kanno2017}.
In the decision-making system based on LFF\cite{Mihana2019}, each slot was assigned to a specific laser, and the player selected the slot corresponding to the leader laser.
By switching the leader--laggard relationship and adjusting the leader probability by using coupling strengths, effective exploration was achieved, and the winning machine was correctly identified.
This decision-making principle is scalable to scenarios with a multitude of slot machine options\cite{Mihana2020}.
However, it is important to note that this system does not realize conflict-free joint decisions. 

To equip this system with the capability of conflict-free joint decision-making, we shift our focus on another synchronous phenomenon known as zero-lag synchronization. 
This phenomenon occurs when the temporal waveforms of semiconductor lasers in a laser network synchronize without any time delay\cite{Nixon2011, Nixon2012, Ohtsubo2015, Mihana2020}.
For example, in a network with three mutually-coupled lasers arranged linearly, the two outer lasers synchronize with zero lag. Meanwhile, an outer laser and an inner laser synchronize with a delay equal to the time required for light to propagate between the coupled lasers\cite{Ingo2006}.
This phenomenon enables lasers that are physically distant from each other to synchronize without delay.
We expect to achieve a decision-making system that is capable of making conflict-free joint decisions using zero-lag synchronization.

In this study, we propose a cooperative decision-making scheme to address the CMAB problem by utilizing a laser network that exhibits both lag synchronization and zero-lag synchronization of LFF dynamics.
Similar to previous work\cite{Mihana2019, Mihana2020}, decisions of each player in our system are made by leveraging the leader--laggard relationships in lasers. Initially, we propose a network for cooperative decision-making and validate its effectiveness through numerical simulations. Subsequently, we implement this network experimentally and demonstrate cooperative decision-making in action.
In these sections, we investigate a basic scenario involving two players and two slot machines (2-player, 2-slot situation), and then we describe the expansion of the system and examine the numerical performance of the expanded system. 
The innovative aspect of this study is realizing the avoidance of selection collision among players by exploiting zero-lag synchronization.
No physical conflict avoidance principle has ever been demonstrated to be so flexible and scalable, and it greatly expands the possibilities for decision-making with photonics.

\section{Results}\label{sec:result}

%=====FIGURE 1=======%
\begin{figure}[t]
\centering
\includegraphics[width=0.9\linewidth]{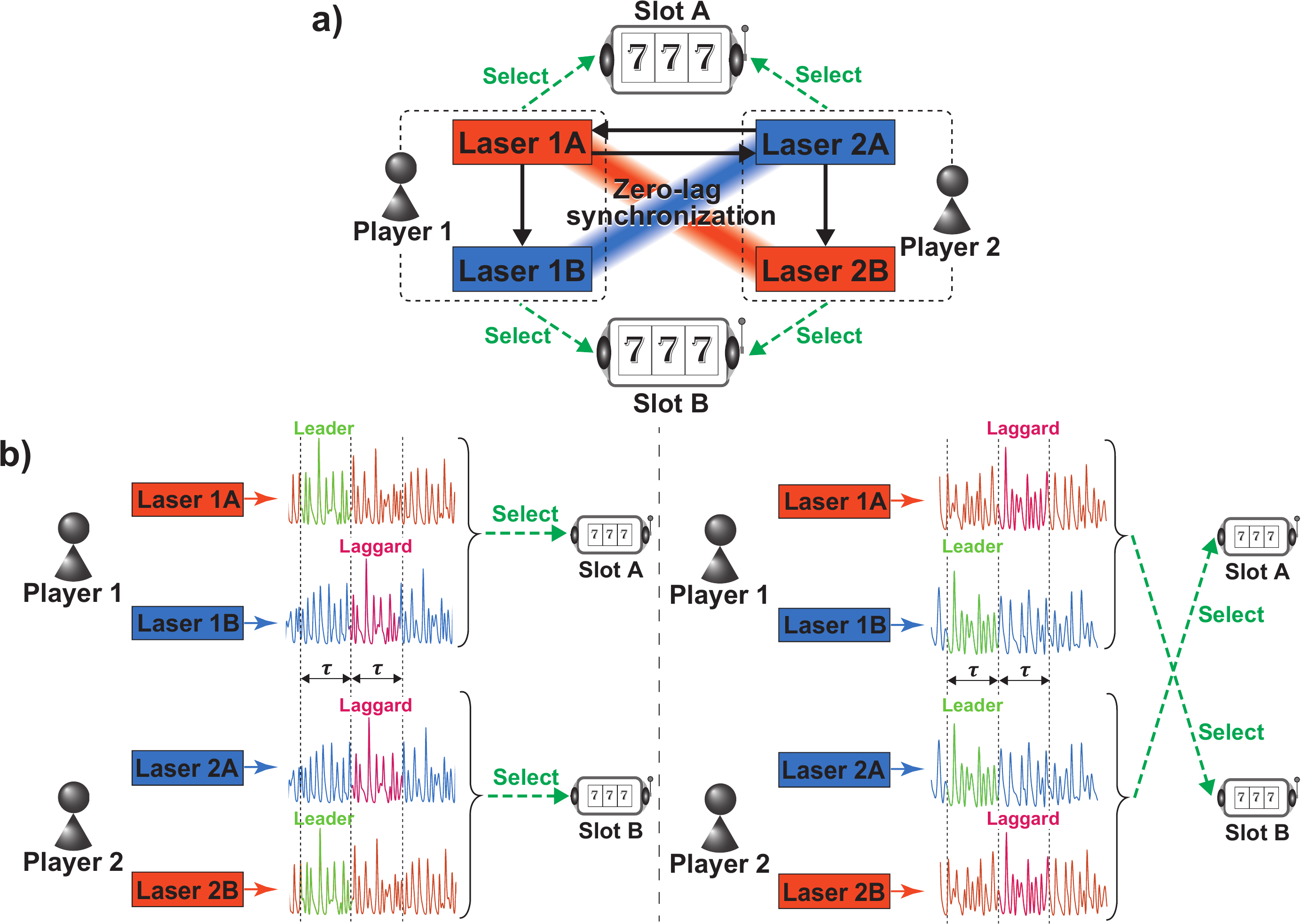}
\caption{System architecture for addressing the competitive multi-armed (CMAB) problem in a 2-player, 2-slot context. 
a) Schematic diagram of decision-making using zero-lag synchronization in a laser network. 
b) Players' decision on slot machine selection facilitated by lag synchronization. The zero-lag synchronization enables conflict-free joint decision-making between Players 1 and 2.}
\label{fig:system}
\end{figure}
%===================%

\subsection{Decision-making method and network configuration}
In this research, we focus on the CMAB problem, which involves multiple players selecting from multiple slots with unknown hit probabilities in order to maximize their profit.
Each slot is assumed to pay a constant reward of 1 according to static probability distributions.
However, if multiple players choose the same slot machine, the rewards are divided among them. 
Consequently, players must solve the CMAB problem using a certain scheme as the MAB problem while avoiding selection conflicts.
First, we consider the decision-making process for two players choosing between two slot machines.
By examining this fundamental situation, we aim to demonstrate a new cooperative decision-making principle.
We believe that this novel principle can be applied to a variety of problems.

Figure \ref{fig:system}a illustrates the principle of decision-making for a 2-player, 2-slot situation using a laser network.
In this system, Lasers 1A and 1B are assigned to Player 1, while Lasers 2A and 2B are assigned to Player 2.
Players make decisions by utilizing the lag synchronization of chaos in two lasers assigned to them.  
At the same time, Lasers 1A and 2A are assigned to Slot A, and Lasers 1B and 2B are assigned to Slot B.
Based on the leader--laggard relationship in lag synchronization of chaos, players select a slot corresponding to the leader laser among the lasers assigned to each of them. 
For instance, if Laser 1A becomes the leader between Lasers 1A and 1B, Player 1 selects Slot A.
Conversely, if Laser 2B becomes the leader between Lasers 2A and 2B, Player 2 selects Slot B.

To achieve cooperative decision-making, zero-lag synchronization in the laser network helps avoid selection collisions among players.
The laser network in Fig. \ref{fig:system}a comprises one set of two mutually-coupled lasers (Lasers 1A and 2A) and two lasers injected from the mutually-coupled lasers (Lasers 1B and 2B).
Lasers 1B and 2B are expected to synchronize with Lasers 2A and 1A, respectively, because Lasers 1B and 2B are injected from Lasers 1A and 2A, respectively, similar to Lasers 2A and 1A.
Theoretically, we can find clusters of lasers in the laser network that synchronize with zero-lag using an adjacency matrix\cite{Nixon2011, Mihana2020}.
In general, each element in an adjacency matrix raised to the $n$-th power provides how many paths of length $n$ from the row to the column are there. 
In addition, by applying the sign function to the power of the adjacency matrix of the laser network, each column in this matrix reveals light injection from itself or other lasers.
The adjacency matrix of laser network ($G$) in Fig. \ref{fig:system}a is as follows:
\begin{align}
\centering
G=\begin{bmatrix}
0&1&1&0\\
0&0&0&0\\
1&0&0&1\\
0&0&0&0\\
\end{bmatrix}.
\end{align}
Each row and column of $G$ represents Laser 1A, 1B, 2A, and 2B in order.
In this case, there are only two types of adjacency matrix $G$ to any $n$-th power with the sign function applied as follows $(k\in N)$:
\begin{align}
\mathrm{sign}(G^n)=
\begin{cases}
\begin{bmatrix}
0&1&1&0\\
0&0&0&0\\
1&0&0&1\\
0&0&0&0\\
\end{bmatrix}
& (n=2k-1),\\
\begin{bmatrix}
1&0&0&1\\
0&0&0&0\\
0&1&1&0\\
0&0&0&0\\
\end{bmatrix}
& (n=2k).
\end{cases}
\end{align}
From this result, Lasers 1A and 2B can synchronize with zero-lag because the elements of the first and fourth columns are the same at any time.
At the same time, Lasers 1B and 2A can synchronize with zero-lag because the elements of the second and third columns are the same.
Therefore, when Player 1 determines that Laser 1A is the leader, Player 2 is expected to determine that Laser 2B is the leader.
Likewise, when Player 1 determines that Laser 1B is the leader, Player 2 is expected to determine that Laser 2A is the leader.
These clusters of zero-lag synchronization help to avoid selection conflicts, as shown in Fig. \ref{fig:system}b.

\subsection{Numerical simulations}
We numerically simulated the performance of this network and the cooperative decision-making.
We modeled the behavior of coupled semiconductor lasers using the Lang-Kobayashi equation as follows\cite{Lang1980} $(j=\mathrm{1A, 1B, 2A, 2B})$:
\begin{align}
\frac{dE_{j}(t)}{dt}&=\frac{1+i\alpha}{2}\left[\frac{G_N\lbrack N_{j}(t)-N_0\rbrack}{1+\epsilon|E_{j}(t)|^2}-\frac{1}{\tau_p}\right] E_{j}(t)+\kappa E_{g(j) }(t-\tau)\exp\lbrack i\theta_{j}(t)\rbrack,\label{eq1}\\
\frac{dN_j(t)}{dt}&=J-\frac{N_j(t)}{\tau_s}-\frac{G_N\lbrack N_j(t)-N_0\rbrack}{1+\epsilon|E_j(t)|^2}|E_j(t)|^2,\\
\theta_j(t)&=(\omega_{g(j)}-\omega_j)t-\omega_{g(j)}\tau,
\end{align}
where $E_j(t)$, $N_j(t)$, and $\theta_j(t)$ are the complex electric-field amplitude, the carrier density, and the optical phase difference between injected light and simplex oscillation light of Laser $j$, at time $t$, respectively.
The second part of Eq. \eqref{eq1} is about the effect of injection light. 
Laser $g(j)$ injects light into Laser $j$.
In this network, $g(j)$ is configured as follows:
\begin{align}
g(j)=
\begin{cases}
\mathrm{2A} & (j=\mathrm{1A}), \\
\mathrm{1A} & (j=\mathrm{1B}), \\
\mathrm{1A}&(j=\mathrm{2A}), \\
\mathrm{2A}&(j=\mathrm{2B}).
\end{cases}
\end{align}
$\kappa$ represents the coupling strength between lasers, and $\tau$ denotes the propagation delay times of an injection path. In this simulation, $\kappa$ is set to \SI{30}{\nano\second^{-1}}, and $\tau$ is set to \SI{5}{\nano\second}.
These factors are critical in understanding synchronization delay and the period of leader--laggard switching, which are important quantities characterizing the system's behavior. 
In this simulation, all detuning at the initial optical frequency is set to \SI{0}{\hertz} to demonstrate the principle of conflict-free operation using the laser network.
Other parameter values used are summarized in Table \ref{table1} in the Methods section.

%=========================%
\begin{figure}[t]
\centering
\includegraphics[width=0.9\linewidth]{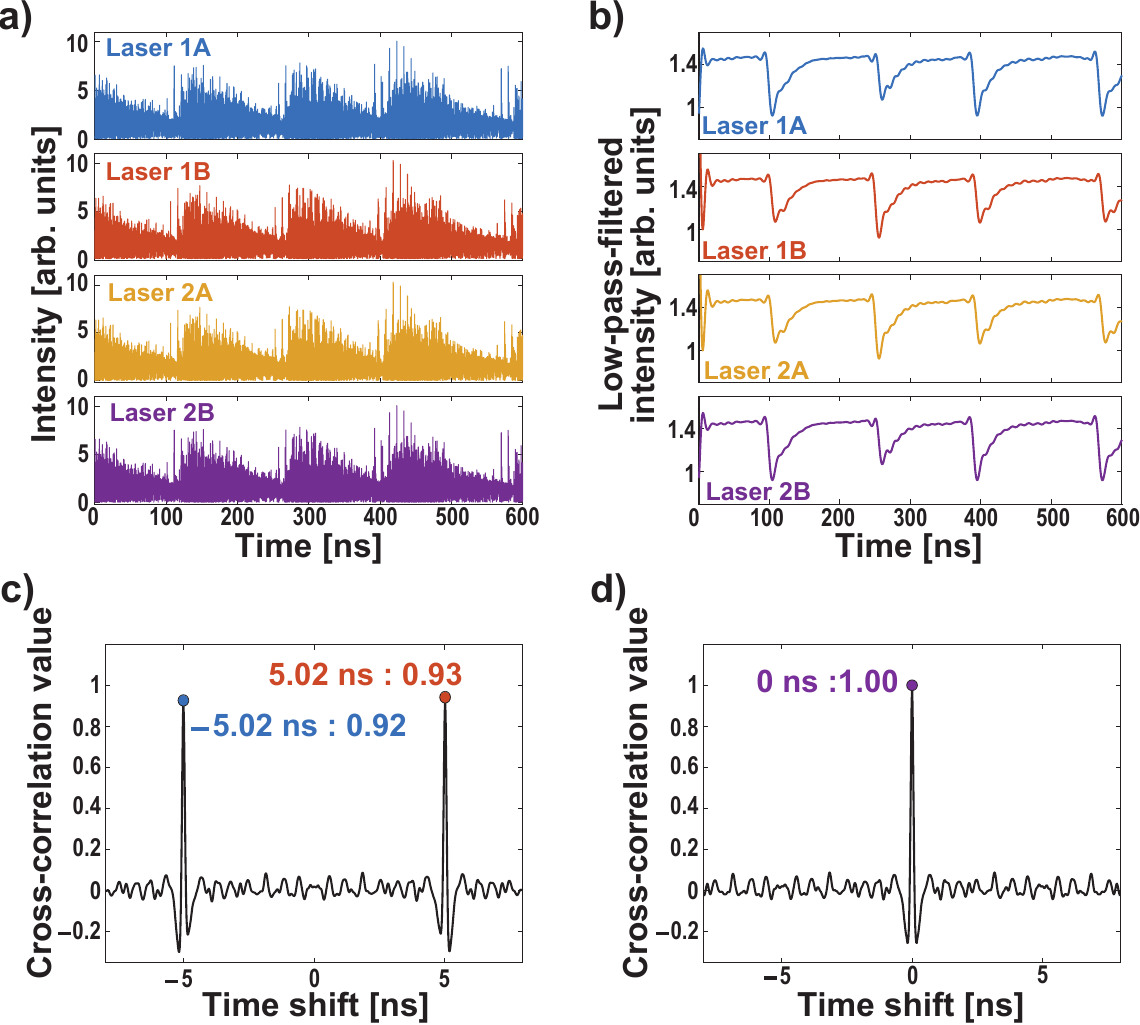}
\caption{Numerical simulation results: about temporal waveforms and cross-correlation values. 
a) Temporal waveforms of laser intensities determined by the Lang-Kobayashi equations. 
b) Low-pass filtered temporal waveforms of laser intensity. 
c) Cross-correlation value between Lasers 1A and Laser 1B. 
d) Cross-correlation value between Lasers 1A and Laser 2B.}
\label{fig:simulation}
\end{figure}
%=========================%

The temporal waveform of laser intensity $I_j(t) = |E_j(t)|^2$ was calculated using the Lang-Kobayashi equation.
Figure \ref{fig:simulation} displays the laser intensity $I_j$, low-pass-filtered intensity, and cross-correlation value between Laser 1A--2B and Laser 1A--1B.
In Fig. \ref{fig:simulation}a, all lasers exhibit chaotic oscillations, but the envelope components present periodic temporal waveforms.
We then examine the low-pass-filtered intensity, as shown in Fig. \ref{fig:simulation}b.
The cutoff frequency is set at \SI{60}{\mega\hertz}. 
Filtered intensities for all lasers demonstrate dropout and gradual recovery, indicating that the lasers oscillate due to LFFs.
Furthermore, their temporal waveforms are strikingly similar.

The cross-correlation value is defined as follows (with $i,j\in \{$1A, 1B, 2A, 2B$\}$):
\begin{align}
    \hat{C}_{i,j}(\tau)=\frac{\langle[I_{i}(t-\tau)-\Bar{I}_{i}][I_{j}(t)-\Bar{I}_{j}]\rangle_T}{\sigma_{i}\sigma_{j}},
\end{align}
where $I_i$ represents the laser intensity of Laser $i$.
Additionally, $\langle\cdot\rangle_T$ denotes the average over the period $T=$\SI{1000}{\nano\second}.
$\bar I_j$ and $\sigma_j$ are the averaged $I_j$ and the standard deviation of $I_j$ over the period $T$, respectively. 
In Fig. \ref{fig:simulation}c, the cross-correlation value $\hat{C}_{1A,1B}$ between Lasers 1A and 1B exhibits peaks at \SI{-5.02}{\nano\second} and \SI{5.02}{\nano\second}, corresponding to the lag synchronization.
Conversely, the cross-correlation value $\hat{C}_{1A,2B}$ between Lasers 1A and 2B has the peak at \SI{0}{\second}, as shown in Fig. \ref{fig:simulation}d, meaning that Lasers 1A and 2B are zero-lag synchronized.
From these results, it is numerically shown that zero-lag synchronization occurs as intended.
%On the other hand, cross-correlation value $\hat{C}_{1A,1B}$ between Lasers 1A and 1B does not have a peak at \SI{0}{\second}, but have two peaks near the propagation delay time $\tau=$\SI{5}{\nano\second}.
%Then, Lasers 1A and 1B occur lag synchronization of chaos.
We also observe that Lasers 2A and 2B exhibit lag synchronization of chaos, while Lasers 1B and 2A are in the zero-lag synchronization. 
Therefore, this laser network allows for the coexistence of lag synchronization chaos and zero-lag synchronization.

We focus on the leader--laggard relationship in the lag synchronization of chaos.
The leader--laggard relationship is quantified using the short-term cross-correlation (STCC) value as follows:
\begin{align}%定義変更済み(fig変更済み)
C_{\mathrm{1A}}(t)&=\frac{\langle[I_{\mathrm{1B}}(t-\tau)-\Bar{I}_{\mathrm{1B}}][I_{\mathrm{1A}}(t)-\Bar{I}_{\mathrm{1A}}]\rangle_\tau}{\sigma_{\mathrm{1A}}\sigma_{\mathrm{1B}}},\\
C_{\mathrm{1B}}(t)&=\frac{\langle[I_{\mathrm{1A}}(t-\tau)-\Bar{I}_{\mathrm{1A}}][I_{\mathrm{1B}}(t)-\Bar{I}_{\mathrm{1B}}]\rangle_\tau}{\sigma_{\mathrm{1A}}\sigma_{\mathrm{1B}}},\\
C_{\mathrm{2A}}(t)&=\frac{\langle[I_{\mathrm{2B}}(t-\tau)-\Bar{I}_{\mathrm{2B}}][I_{\mathrm{2A}}(t)-\Bar{I}_{\mathrm{2A}}]\rangle_\tau}{\sigma_{\mathrm{2A}}\sigma_{\mathrm{2B}}},\\
C_{\mathrm{2B}}(t)&=\frac{\langle[I_{\mathrm{2A}}(t-\tau)-\Bar{I}_{\mathrm{2A}}][I_{\mathrm{2B}}(t)-\Bar{I}_{\mathrm{2B}}]\rangle_\tau}{\sigma_{\mathrm{2A}}\sigma_{\mathrm{2B}}},
\end{align}
%=====FIG 3============================%
\begin{figure}[tb]
\centering
\includegraphics[width=0.9\linewidth]{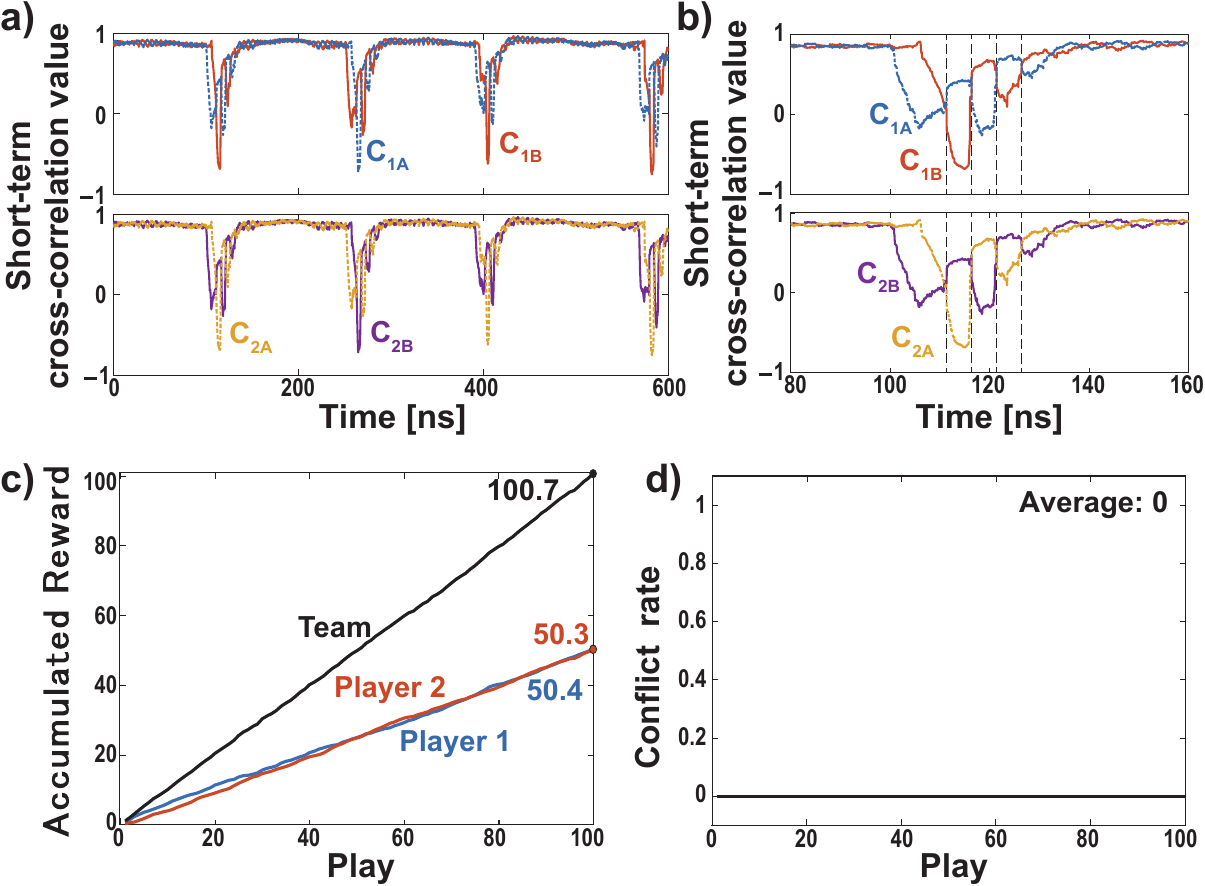}
\caption{Numerical simulation results: short-term cross-correlation values and decision-making. 
a) Short-term cross-correlation values $C_{1A}$ (blue curve), $C_{1B}$ (orange curve), $C_{2A}$ (yellow curve), and $C_{2B}$ (purple curve) calculated from Fig. \ref{fig:simulation}a. 
b) Enlarged view of short-term cross-correlation values. 
c) Accumulated rewards for Player 1 (blue curve), Player 2 (orange curve), and the team (black curve) in the decision-making. 
d) Conflict rate between Players 1 and 2 in decision-making.}
\label{fig:simulation2}
\end{figure}
%=====================================%
where $\langle\cdot\rangle_\tau$ denotes the short-term average over the period $\tau$.
$\bar I_j$ and $\sigma_j$ represent the averaged $I_j$ and the standard deviation of $I_j$ over the period $\tau$, respectively.
Note that the short-term cross-correlation value has time $t$ as a parameter, rather than the delay time $\tau$, which is used in the cross-correlation value.
By employing STCC values, we can observe the changes in cross-correlation values between lasers assigned to each player as a function of time.
$C_\mathrm{1A}(t)$ represents the cross-correlation value when Laser 1A is considered as the laggard over the short-term period $\tau$, while $C_\mathrm{1B}(t)$ considers Laser 1B as the laggard. 
Consequently, at time $t$, if $C_\mathrm{1A}$ is smaller than $C_\mathrm{1B}$, Laser 1A is the leader and Laser 1B is the laggard; the reverse is also true.
Similarly, $C_{2A}$ and $C_{2B}$ reveal the relationships between Lasers 2A and 2B.

Using this temporal waveform in Fig \ref{fig:simulation}a, we calculate the STCC values and perform decision-making numerically. Figure \ref{fig:simulation2} displays the STCC values and decision-making results.
Comparing Figs. \ref{fig:simulation}b and \ref{fig:simulation2}a, it can be observed that the LPF and STCC correspond closely, particularly during the dropout and recovery processes.
Figure \ref{fig:simulation2}b presents an enlarged view of Fig \ref{fig:simulation2}a.
Despite the one-way coupling between Lasers 1A and 1B, STCC values $C_{1A}$ and $C_{1B}$ spontaneously switch every \SI{5}{\nano\second}, corresponding to the coupling delay time, as seen in lag synchronization of chaos.
Moreover, STCC values $C_{1A}$ and $C_{1B}$ are the same as $C_{2B}$ and $C_{2A}$, respectively, since each laser is synchronized with zero-lag (Laser 1A--2B and Laser 1B--2A).
Decision-making is performed using the switching of these correlation values.
The sampling interval for players' selecting slot machines is set to \SI{1}{\nano\second}, equivalent to \SI{1}{play}.
We expect that the selection of each player will switch without selection conflict every \SI{5}{plays} because the coupling delay time is set to \SI{5}{\nano\second}.
The hit probabilities of Slots A and B are set to 0.8 and 0.2, respectively.
We test the cooperative decision-making for \SI{100}{plays}.
We define \SI{100}{plays} as one cycle and evaluate the average accumulated reward for \SI{10}{cycles}.
Figure \ref{fig:simulation2}c displays the averaged accumulated reward, while Fig. \ref{fig:simulation2}d displays the averaged decision conflict rate over \SI{10}{cycles}.
In this problem, the team's average theoretical maximum reward of 100 can be achieved if no selection conflict occurs.
In Fig. \ref{fig:simulation2}c, the accumulated rewards of Players 1 and 2 are 50.4 and 50.3, and that of the team is 100.7.
Also, no conflict occurs in 10 cycles, as shown in Fig. \ref{fig:simulation2}c.
Thus, using this network, cooperative decision-making is successfully achieved, and the reward for individuals and the team reaches the theoretical maximum reward.
In addition, players are rewarded equally, even though the hit probabilities of slots are biased. 
This equality is attained because the selections of players are constantly switched, providing equal opportunities for players to choose the best slot. 

%====FIG 4=========================%
\begin{figure}[t]
\centering
\includegraphics[width=0.8\linewidth]{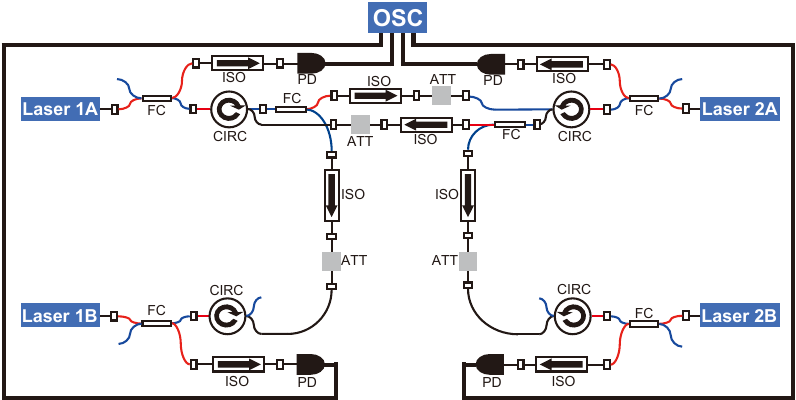}
\caption{Experimental setup of the proposed decision-making system. 
ATT: voltage-controlled variable attenuator, 
CIRC: optical circulator, 
ISO: optical isolator, 
FC: fiber coupler, 
OSC: oscilloscope, 
PD: photodetector.}
\label{fig:setup}
\end{figure}
%=================================%
\subsection{Experimental results}
\subsubsection{Conflict-free joint decision by two players}
We implement the experimental setup for the 2-player, 2-slot situation as shown in Fig. \ref{fig:setup} to confirm the experimental possibility of our system for this fundamental situation.
We use four distributed-feedback semiconductor lasers, referred to as Laser 1A, Laser 1B, Laser 2A, and Laser 2B (threshold injection current values are \SIlist{11.0;11.7;11.8;11.8}{\milli \ampere}, respectively.).
The light from each laser is split into a network part and a detection part by a fiber coupler.
In the detection part, the light is transferred to an electrical signal by a photoreceiver and sent to the oscilloscope.
In the network part, the four lasers are connected via unidirectional injection paths with optical isolators, fiber couplers, and optical circulators to configure the network in Fig. \ref{fig:setup}.
Each path is adjusted to have the same delay time of \SI{75.37}{\nano\second}. 
In addition, a voltage-controlled variable attenuator is inserted to adjust the coupling strength $\kappa$.
$\kappa_{i,j}$ is the coupling strength of the injection path from Laser $i$ to Laser $j$.
Circulators are also inserted on the Laser $k$B side to make the lengths and the coupling strengths of four injection paths the same ($k=1, 2$). 
In order to achieve lag synchronization of chaos in the LFF dynamics, the injection current of each laser is set to about 1.1 times the threshold injection current, and the temperature is set so that the peak wavelength is \SI{1547.100}{\nano\metre}.
In this network, mutually coupled Lasers 1A and 2A cause LFF dynamics. 
Lasers 1A and 2A are also transmitted to Lasers 1B and 2B, respectively, with the delay,  Lasers 1A and 2B (1B and 2A) synchronize without delay.
Therefore, the coupling strengths between Lasers 1A and 2A are set to $\kappa_{1A,2A}=0.67$ and $\kappa_{2A,1A}=0.73$ to oscillate in LFF dynamics.
On the other hand, the coupling strengths between Lasers 1B and 2B are set to $\kappa_{1A,1B}$ and $\kappa_{2A,2B}=1.0$ to achieve the lag synchronization.
%====FIG 5 ========================%
\begin{figure}[t]
\centering
\includegraphics[width=0.9\linewidth]{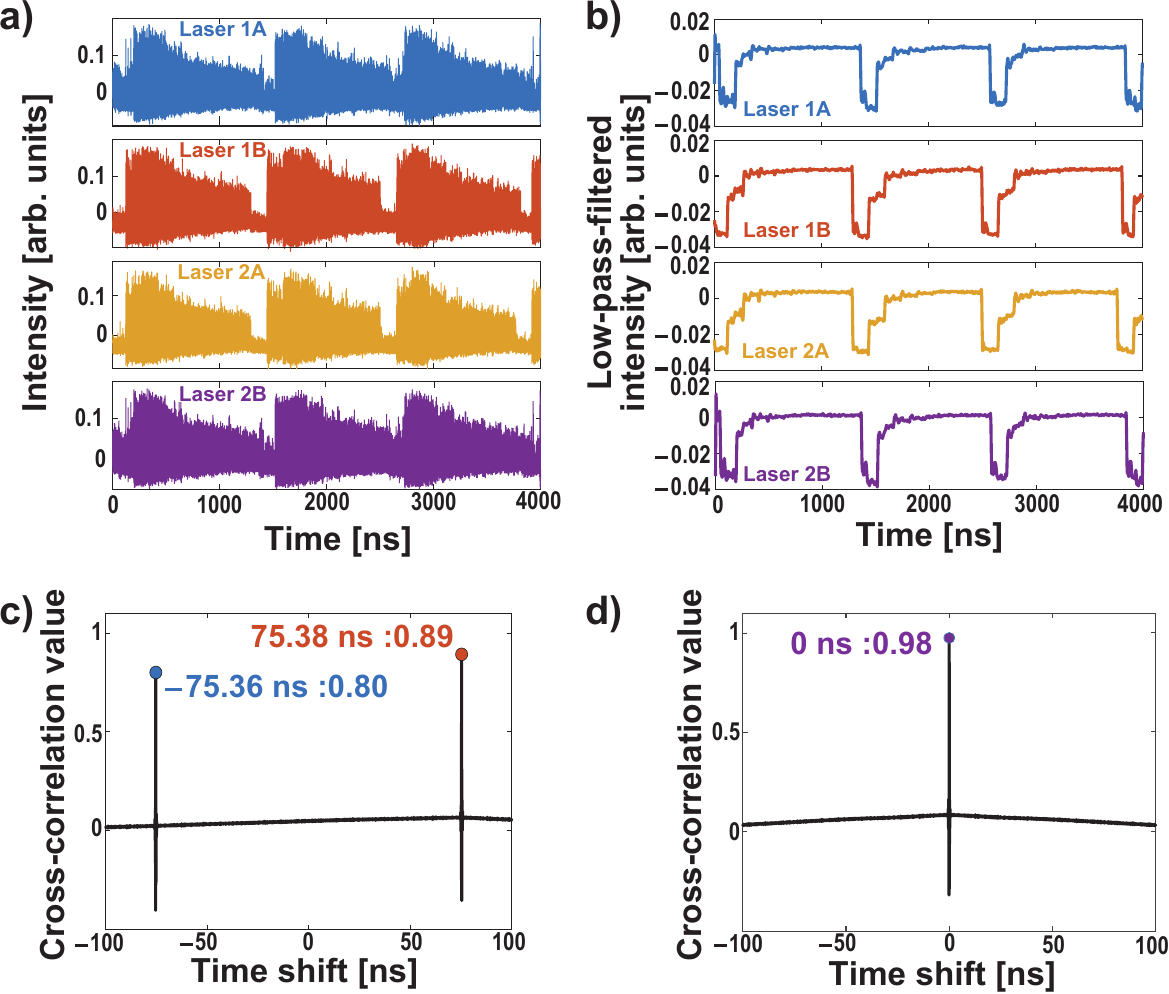}
\caption{Experimental results: temporal waveforms and cross-correlation values. 
a) Temporal waveforms of laser intensities. 
b) Low-pass filtered temporal waveforms of laser intensities. 
c) Cross-correlation value between Lasers 1A and Laser 1B ($\hat C_{1A,1B}(0)$). 
d) Cross-correlation value between Lasers 1A and Laser 2B ($\hat C_{1A,2B}(0)$). }
\label{fig:expwaveform}
\end{figure}
%=================================%

%====FIG 6 ========================%
\begin{figure}[t]
\centering
\includegraphics[width=0.9\linewidth]{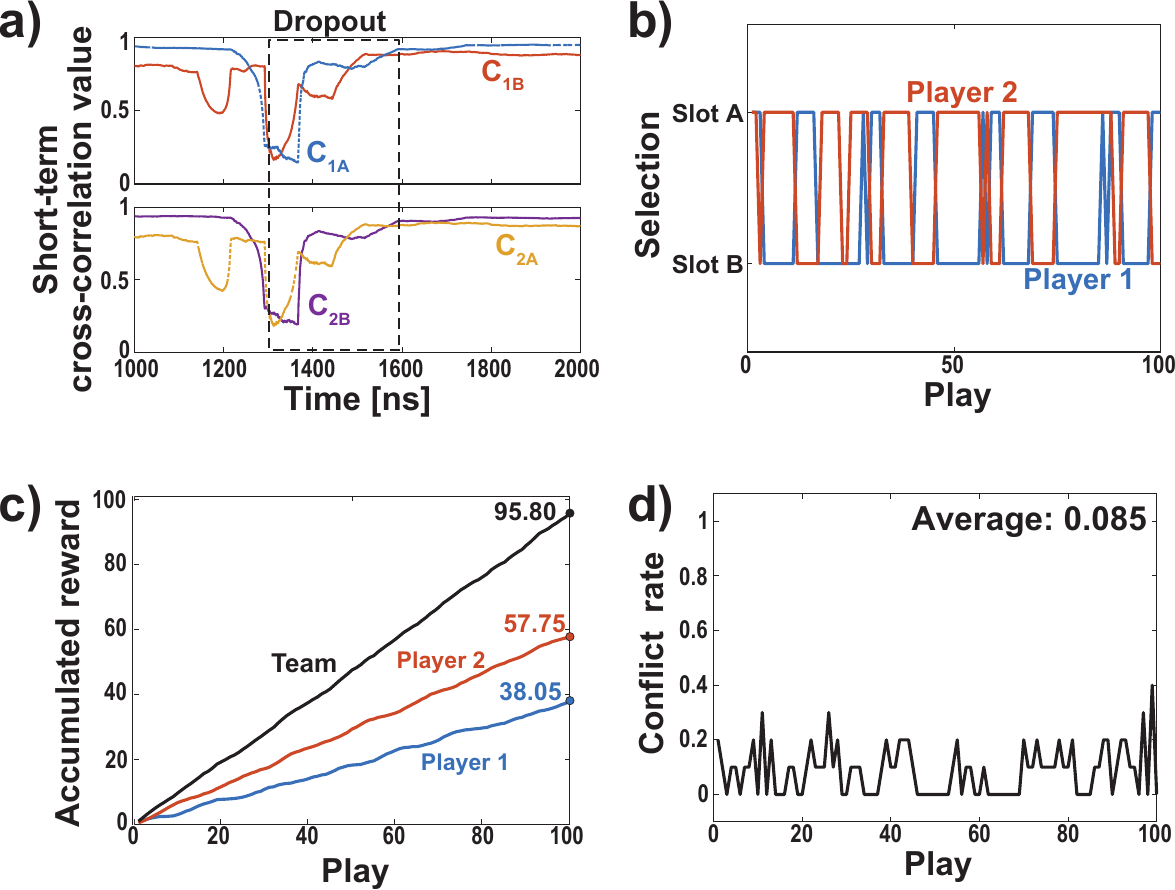}
\caption{Experimental results: short-term cross-correlation values and decision-making. 
a) Enlarged view of short-term
cross-correlation values $C_{1A}$ (blue curve), $C_{1B}$ (orange curve), $C_{2A}$ (yellow curve), and $C_{2B}$ (purple curve) calculated from Fig.\ref{fig:expwaveform}a. 
b) Selection of Player 1 (blue curve), Player 2 (orange curve), and the team (black curve) in the decision-making. 
c) Averaged accumulated reward of Player 1 (blue curve), Player 2 (orange curve), and team (black curve) in the decision-making. 
d) Averaged conflict rate between Player 1 and 2 in the decision-making.}
\label{fig:expdecisionmaking}
\end{figure}
%=================================%

Figure \ref{fig:expwaveform} displays the waveforms and cross-correlation (CC) values for the laser intensities in the experiment.
Figure \ref{fig:expwaveform}a shows the temporal waveforms, showing that the lasers oscillate chaotically with the long periods.
In this figure, Laser 1A (1B) is similar to Laser 2B (2A) in the range of small oscillation amplitudes.
Figure \ref{fig:expwaveform}b shows the low-pass-filtered temporal waveforms.
The cutoff frequency of the low-pass filter is \SI{60}{\mega\hertz}.
Comparing this with Fig. \ref{fig:expwaveform}a, Laser 1A (1B) clearly synchronizes with Laser 2B (2A).
Then, we check the cross-correlation values of Lasers 1A--1B (Fig. \ref{fig:expwaveform}c) and Lasers 1A--2B (Fig. \ref{fig:expwaveform}d).
In Fig. \ref{fig:expwaveform}c, there are two peaks near \SI{75.37}{\nano\second}, corresponding to the propagation delay time of the experimental setup. 
In fact, there are differences in the actual lengths of the injection paths, and the difference of the propagation delay time is within approximately \SI{1}{\nano\second}. 
To compensate for these differences, the time positions of waveforms are adjusted using the oscilloscope's deskew feature. Details of how the deskew is adjusted are described in previous research\cite{Mihana2020}.
Although the positions of peaks are adjusted near \SI{75.37}{\nano\second}, there are still a few differences between delays because of the resolution of the oscilloscope. 
In Fig. \ref{fig:expwaveform}d, the CC value reaches 0.98 in the case of no delay.
Then, Lasers 1A and 2B are regarded as achieving zero-lag synchronization.

Using this temporal waveform in Fig. \ref{fig:expwaveform}a, we calculate the STCC values and perform a decision-making experiment.
Figure \ref{fig:expdecisionmaking}a shows the STCC values.
Unlike the case of numerical calculation, the switching between the STCC values occurs in the dropout part.
However, leader--laggard switching is crucial to achieve equal rewards among players and effective exploration.
Thus, in the experimental system, we focuses on regions near the dropout where switching took place.
As shown in Fig. \ref{fig:expdecisionmaking}a, dropout is determined based on STCC values.
Decision-making begins when all of the STCC values $C_{1A}$, $C_{1B}$, $C_{2A}$, and $C_{2B}$ drop below 0.4 and stops when any one of the STCC values rises above 0.94.
The decision-making is made every \SI{15.06}{\nano\second}, \SI{10}{cycles} each consisting of \SI{100}{plays} are performed, similar to the numerical simulation. 
Since the propagation delay time is about five times the decision-making period, player selections are expected to switch every \SI{5}{plays}.
For this decision-making experiment, a waveform of approximately \SI{15}{\micro\second} is needed.%%%
Figure \ref{fig:expdecisionmaking}b displays the selection of each player in one cycle, while Fig. \ref{fig:expdecisionmaking}c presents the averaged accumulated rewards of each player and the team. 
Figure \ref{fig:expdecisionmaking}d demonstrates the averaged conflict rate between players in \SI{10}{cycles}.
In Fig. \ref{fig:expdecisionmaking}b, players mostly select different slot machines. 
Then, the accumulated rewards of Player 1, Player 2, and the team are 38.05, 57.75, and 95.8, respectively, and the average conflict rate is 0.085.
We can see that the experimental results support the results of numerical simulations; however, there is a selection conflict, and the reward is reduced accordingly.
This is because the experiments include noise, and zero-lag synchronization is not perfectly achieved.
Based on these results, the cooperative decision-making system can successfully reduce the conflict rate and increase rewards for the players and the team numerically and experimentally. 

%=====FIG 7 ======================%
\begin{figure}[t]
\centering
\includegraphics[width=0.9\linewidth]{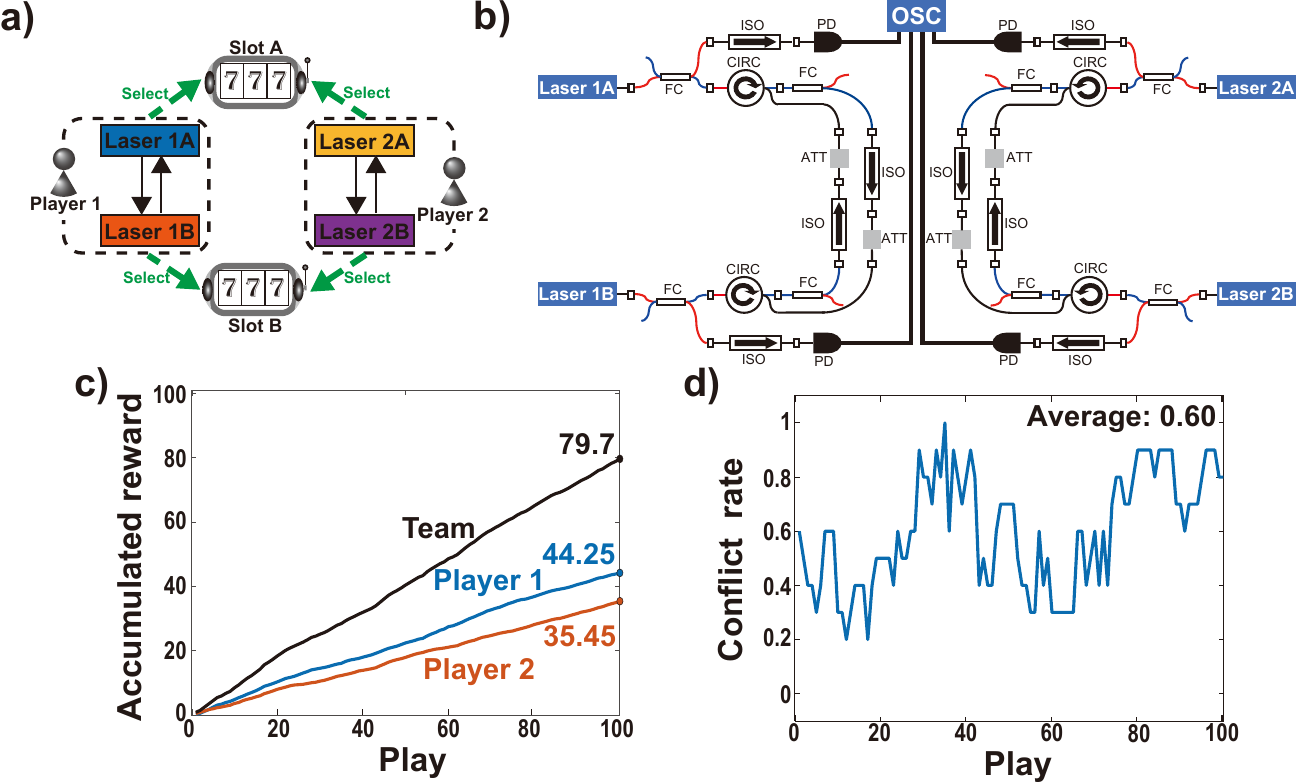}
\caption{Comparative experiment description and result. 
a) Schematic diagram illustrating decision-making using the two independent pairs of mutually coupled lasers. 
b) Experimental setup of the comparative experiment. 
c) Averaged accumulated rewards of Player 1 (blue curve), Player 2 (orange curve), and team (black curve) during the decision-making process. 
d) Averaged conflict rate in the decision-making.}
\label{fig:indepexp}
\end{figure}
%=================================%

\subsubsection{Decision-making by independent two-players}
As a comparative experiment, the decision-making system is tested using two pairs of independently mutually coupled lasers, as shown in Figs. \ref{fig:indepexp}a and b. 
Laser 1A--1B and Laser 2A--2B are mutually coupled independently through unidirectional injection paths. 
Player 1 is assigned Lasers 1A and 1B, while Player 2 is assigned Lasers 2A and 2B. 
Lasers 1A and 2A are assigned to Slot A, and Lasers 1B and 2B are assigned to Slot B, similar to the system in Fig. \ref{fig:system}a. 
This setup represents a situation where players make decisions independently.
As shown in Fig. \ref{fig:indepexp}b, each injection path is composed of circulators, a fiber coupler, an attenuator, and an isolator. 
The fiber couplers are inserted to match the injection path lengths to the experimental setup in Fig. \ref{fig:setup}. 
The coupling strengths for the mutually coupled lasers, $\kappa_{1A,1B}$, $\kappa_{1B,1A}$, $\kappa_{2A,2B}$, and $\kappa_{2B,2A}$ are set to \SIlist{0.339;0.97;0.96;0.67}{}, respectively.

In the comparative experiment with independent decision-making, the coupling strengths are set to generate LFF dynamics, and 
the biased pairs of coupling strengths compensate for the differences in the characteristics of semiconductor lasers to achieve balanced leader probability for each laser. 
In this scenario, the temporal waveforms in LFFs of the two lasers assigned to each player are synchronized with a delay equal to the propagation delay time, but there is no observed correlation between the lasers assigned to Player 1 and those assigned to Player 2. 
Using the delay synchronization of chaos, the decision-making experiment is performed similarly to the previous cooperative decision-making experiment.
Figure \ref{fig:indepexp}c displays averaged accumulated rewards for each player and the team, and Fig. \ref{fig:indepexp}d presents the averaged conflict rate between players across \SI{10}{cycles}. 
The averaged accumulated rewards of Player 1, Player 2, and the team are 35.45, 44.25, and 79.7, respectively. 
The averaged conflict rate is 0.60. 
Comparing the results of the independent decision-making system with those of the cooperative decision-making system, it becomes evident that the cooperative approach outperforms the independent one. 
In particular, zero-lag synchronization functions as a conflict-free joint decision in the decision-making system.
The cooperative decision-making system achieves a higher team reward and a significantly lower conflict rate, highlighting its effectiveness in improving decision-making processes and outcomes.

\section{Discussion}
%スイッチングについて理論との差異 
%具体的にどんな応用があるか
% The Discussion should be succinct and must not contain subheadings.
\subsection{Performance and stability}
In these results, numerical and experimental results are similar and are superior to the independent system. 
However, there is still a gap between the numerical and the experimental result of our decision-making system.
First, perfect conflict-free decision-making is not achieved in the experimental setup. 
Although the achieved conflict rate was 0 in the numerical simulation, the averaged conflict rate in the experimental setup is about \SI{8.5}{\%}.
Second,  perfect equality is not achieved in the experimental setup. 
It is difficult in the experiment to observe the ideal switching of the leader--laggard relationship seen in a numerical simulation.
In the experimental setup, the switching occurs only around dropout, and the leader probability is biased.
Therefore, using the whole waveform similar to the numerical simulation makes the reward of players biased.
Even when we use near the dropout to try to achieve uniform selection switching, the averaged accumulated rewards of Player 1 and Player 2 are 38.0 and 57.7, respectively, making it difficult to guarantee adequate equality.

These deviations between the numerical and experimental results can be attributed to small differences in parameters.
In the numerical simulation, the propagation delay time $\tau$ of all injection paths and many parameters of semiconductor lasers comprising the network are precisely the same. 
This matching of parameters makes the synchronous state more stable and the system performance more ideal.
Allowing small differences in these parameters, perfect conflict avoidance cannot be achieved, and a few conflicts may occur.
The deterioration of the synchronization status can also make the lasers' dynamics unstable.
Some degree of misalignment in the inner parameters of lasers can be compensated by adjusting the injection current and temperature of the lasers or the coupling strength of injection paths.
Also, these parameters are devised and determined in this experiment to obtain better results.
However, it is presumably difficult to compensate for the differences of $\tau$ in each path by adjusting other parameters. 
In other words, misalignment of $\tau$ among injection paths presumably works critically on the performance and stability of the system because the misalignment in $\tau$ causes a misalignment of arriving light to the lasers that should be synchronized. 
Then, using elaborate delay lines in the adjustment of $\tau$, although this experiment is conducted with deskew, can improve performance and stability.  

Additionally, other networks can achieve the same synchronous state as our network. 
For example, a four-laser network with connections between any adjacent pairs of lasers\cite{Ohtsubo2015} can theoretically and numerically perform just as well as our network. 
However, it becomes increasingly challenging in the experiment to match the $\tau$ of all injection couplings as the number of coupling increases. 
Thus the number of injection couplings should be minimized. 
Our network has the minimum number of injection couplings. 
It can be expanded to situations with more players or slots while maintaining the minimum number of injection couplings.

\subsection{Scalability}
Our network can be expanded to accommodate more players and more slots. 
An adjacency matrix corresponding to the situation that has the same number of players and slots ($n$ players and $n$ slots) is the $n^2\times n^2$ matrix as shown below:
\begin{align}
\centering
G=\begin{bmatrix}
A&O&O&&O&B\\
B&A&O&\dots&O&O\\
O&B&A&&O&O\\
&\vdots&&\ddots&\vdots\\
O&O&O&\dots&A&O\\
O&O&O&\dots&B&A\\
\end{bmatrix}
\end{align}
where the matrixes $A$ and $B$ are the $n\times n$ matrixes as shown below:
\begin{align}
\centering
A=\begin{bmatrix}
0&1&0&\dots&0\\
0&0&1&&0\\
&\vdots&&\ddots&\vdots\\
0&0&0&\dots&1\\
0&0&0&&0\\
\end{bmatrix};
B=\begin{bmatrix}
1&0&0&\dots&0\\
0&0&0&&0\\
&\vdots&&\ddots&\vdots\\
0&0&0&\dots&0\\
0&0&0&&0\\
\end{bmatrix}.
\end{align}
Similarly, the network can be expanded for different numbers of players and slots.
%========FIG 8====================%
\begin{figure}[tb]
\centering
\includegraphics[width=0.9\linewidth]{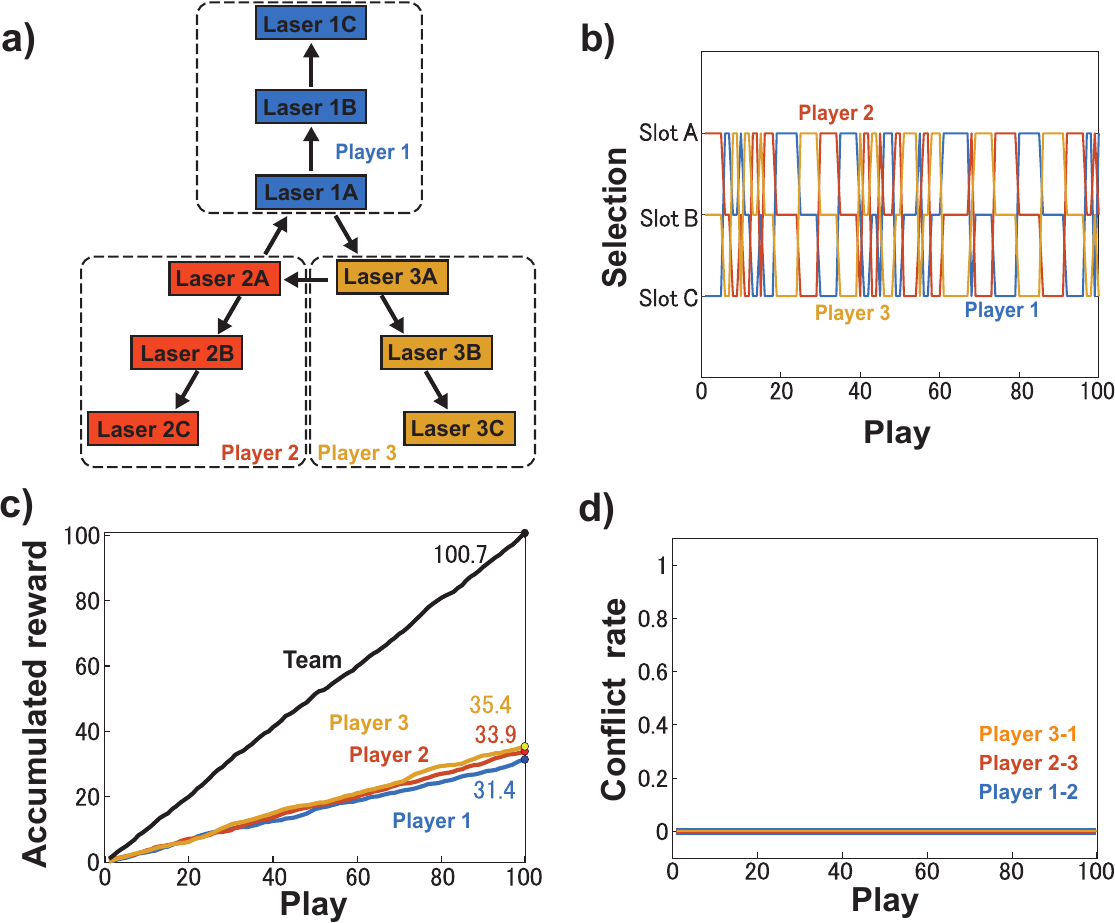}
\caption{Scalability investigation. 
a) Network configuration of the decision-making for three-player, three-slot  situation.  
b) Selections of Player 1 (blue curve), Player 2 (red curve), and Player 3 (orange curve) in the decision-making. 
c) Averaged accumulated rewards of Player 1 (blue curve), Player 2 (red curve), Player 3 (orange curve), and team (black curve) in the decision-making. 
d) Averaged conflict rate between Players 1 and 2 (blue curve), Players 2 and 3 (red curve), and Players 3 and 1 (orange curve) in the decision-making.}
\label{fig:expansion}
\end{figure}
%=================================%
As an example of expanding the situation, the network corresponding to the expanded three-player, three-slot situation is shown in Fig. \ref{fig:expansion}a.
We simulate the decision-making for this expanded situation numerically.
In this network, Laser $j$A, Laser $j$B, and Laser $j$C are assigned to Player $j$ ($j=1, 2, 3$), and Laser 1$k$, Laser 2$k$, and Laser3$k$ are assigned to Slot $k$ ($k=$A, B, C).
In this network, the selection of each player would be decided based on leader--laggard relationships among each player's three lasers, like a ring configuration.
Decision-making by a laser network in a ring configuration has already been established\cite{Mihana2020}. Referring to this, the STCC in lasers assigned to each player is defined similarly to Eqs. \eqref{defstcc1}--\eqref{defstcc3} as follows $(j=1, 2, 3)$:
\begin{align}
C_{j\mathrm{A}}(t)&=\frac{\langle[I_{j\mathrm{C}}(t-\tau)-\Bar{I}_{j\mathrm{C}}][I_{j\mathrm{A}}(t)-\Bar{I}_{j\mathrm{A}}]\rangle_\tau}{\sigma_{j\mathrm{C}}\sigma_{j\mathrm{A}}},\label{defstcc1}\\
C_{j\mathrm{B}}(t)&=\frac{\langle[I_{j\mathrm{A}}(t-\tau)-\Bar{I}_{j\mathrm{A}}][I_{j\mathrm{B}}(t)-\Bar{I}_{j\mathrm{B}}]\rangle_\tau}{\sigma_{j\mathrm{A}}\sigma_{j\mathrm{B}}},\label{defstcc2}\\
C_{j\mathrm{C}}(t)&=\frac{\langle[I_{j\mathrm{B}}(t-\tau)-\Bar{I}_{j\mathrm{B}}][I_{j\mathrm{C}}(t)-\Bar{I}_{j\mathrm{C}}]\rangle_\tau}{\sigma_{j\mathrm{B}}\sigma_{j\mathrm{C}}}.
\label{defstcc3}
\end{align}
$C_{jk} (k=\mathrm{A}, \mathrm{B}, \mathrm{C})$ represents the cross-correlation value when Laser $jk$ is considered as the laggard over the short-term period $\tau$, for example, $C_\mathrm{1B}(t)$ considers Laser 1B lagging behind Laser 1A.
Consequently, among lasers assigned to each player, the one having the minimum STCC value is the leader laser.
For example, if $C_\mathrm{1A}$ is smaller than $C_\mathrm{1B}$ and $C_\mathrm{1C}$, Laser 1A is the leader among Lasers 1A, 1B, and 1C, and other cases are also true.

Parameter values used are summarized in Table \ref{table1} in the Methods section.
Players' selections are every \SI{1}{\nano\second}, which is the same as the numerical simulation, corresponding to 5 identical selections in the duration of the switching of the leader.
In each cycle, Slots A, B, and C payout rewards of 1 with probabilities of 0.6, 0.3, and 0.1, respectively.
Figure \ref{fig:expansion}b displays the selection made by each player in one cycle, while Fig. \ref{fig:expansion}c displays the averaged accumulated reward of each player and team, and Fig. \ref{fig:expansion}d displays the averaged conflict rate between players over 10cycles.
The conflict among the three players is completely avoided, and team or player rewards are maximized.
By switching selections of players, each player receives rewards almost equally, although the probabilities of slots are biased. 
If three players make decisions independently, the expected conflict rate is about \SI{77}{\%}, and the expected rewards of the team and each player are about 56 and 18, respectively. 
The expected conflict rate would grow, and the expected rewards of the team would decrease with independent selection as the number of players increases.
Thus, the effect of this conflict-free principle is also more significant in expanded situations.

Thus, this system works well in extended situations.
As mentioned earlier, it is possible to implement this expanded system experimentally by devising a way to align the length of each injection coupling.

\section{Conclusion}
In this research, we have proposed a cooperative decision-making system and have confirmed its applicability. 
In the numerical simulations, the averaged accumulated rewards of Player 1, Player 2, and the team are 50.4, 50.3, and 100.7, respectively.
When the players make decisions fully cooperatively with no conflict, the expected reward for each player is 50, and for the team is 100.
The numerical simulation achieved this theoretical maximum reward.
When players make decisions independently, the expected reward for each player is 37.5, for the team is 75, and the expected conflict rate is 0.5. The conflict rate and reward improved significantly thanks to the proposed cooperative mechanism. This demonstrates the usefulness of the proposed system numerically.
In the experimental setup, the averaged accumulated rewards of Player 1, Player 2, and the team are 38.0, 57.7, and 95.8, respectively, with an average conflict rate of 0.085. 
In the comparative experiment with independently coupled lasers corresponding to the non-cooperative decision-making, the averaged accumulated rewards of Player 1, Player 2, and the team are 35.4, 44.2, and 79.7, respectively, with an average conflict rate of 0.60. 
These results confirm the superiority of our system.
Furthermore, this system can be expanded to situations with more players and slots.  Network expansion is theoretically generalized. In the three-player, three-armed situation, perfect conflict-free decision-making is also achieved numerically, and the averaged accumulated rewards of Player 1, Player 2, Player 3, and the team are 31.4, 33.9, 35.4, and 100.7, respectively.

There are still issues that need to be resolved to fully realize our system's numerical and theoretical excellent performance in the experimental setup.
All propagation delay times of injection paths presumably should be matched more than the current situation for experimental performance and stability. 
We believe that this issue can be solved technically, and we expect that this system will be further expanded and realized in the experimental system in the future.

A series of studies on decision-making in laser networks has special significance in research on chaotic lasers.
In many examples of applied research so far, the chaotic laser has played a role similar to a signal generator, such as in physical random number generation\cite{Uchida2008} and secure key generation\cite{Koizumi2013}, or a converter, such as in reservoir computing\cite{Larger2012}.
Therefore, the chaotic laser system was insensitive to the application in both system and dynamics.
However, in decision-making using laser networks\cite{Mihana2019, Mihana2020} and multimode lasers\cite{Iwami2022}, the laser system can be directly changed by the result of decision-making.
In other words, the laser system will be applied to a decision-making environment.
On the other hand, photonic decision-making seems to be one of the applications of controlling chaos\cite{ott1990} because decision-making controls laser systems.
We hope these decision-making studies will reboot research related to controlling chaos, and usher in the dawn of adaptive physical systems in not only chaotic laser but also light research.

\section{Methods}
\subsection{Details of numerical experiments}
The parameters of the numerical simulation are described in Table \ref{table1}.
The parameters are determined based on references\cite{Kanno2017,Mihana2019, Mihana2020}.
$\Delta f_{\rm ini}$ refers to the initial optical frequency detuning of each laser from the frequency of Laser 1A, 
$f_{1A}=c/\lambda_{1A}$.
\begin{table}[ht]
 \centering
 \caption{Parameters of the numerical simulation}
  \begin{tabular}{|l|c|c|}
   \hline 
   Symbol&Parameter&Value\\ \hline \hline
   $G_N$&Gain coefficient&\SI{8.40e-13}{\metre^3\second^{-1}}\\ \hline 
   $N_0$&Carrier density at transparency&\SI{1.40e24}{\metre^3}\\ \hline
   $\epsilon$&Gain saturation coefficient&\SI{2.0e-23}{}\\ \hline
   $\tau_p$&Photon lifetime&\SI{1.927e-12}{\second}\\ \hline
   $\tau_s$&Carrier lifetime&\SI{2.04e-9}{\second}\\ \hline
   $\alpha$&Linewidth enhancement factor&\SI{3.0}{}\\ \hline
   $\lambda_{1A}$&Optical wavelength of laser 1A&\SI{1.537e-6}{\metre}\\ \hline
   $c$&Speed of light&\SI{2.998e8}{\metre\second^{-1}}\\ \hline
   $\kappa$&Coupling strength&\SI{30.00e-9}{\second^{-1}}\\ \hline
   $j=J/J_{th}$&Normalized injection current&\SI{1.1}{}\\ \hline
   $\tau$&Propagation delay time of light&\SI{5.0e-9}{\second}\\ \hline
   $\Delta f_{\mathrm{ini}}$&Initial optical frequency detuning&\SI{0}{\hertz}\\
   \hline 
  \end{tabular}
 \label{table1}
\end{table}
\subsection{Details of experiments}
Details of the experimental equipment are described in Table \ref{table2}. In this research, semiconductor lasers have no isolator. This feature allows light to be injected into the laser from another laser. 
In our network, two types of fiber couplers are used, with light split ratios of 90:10 and 50:50. The coupler adjacent to each laser has a 90:10 split, and \SI{10}{\%} of the lasers output is sent to the oscilloscope via an isolator and a photoreceiver. The couplers coordinating the injection paths have a 50:50 split. The type of semiconductor laser is a distributed-feedback semiconductor laser. The threshold injection current value and actual injection current value of each laser used in the experiment are shown in Table \ref{table3}.

\begin{table}[ht]
 \centering
 \caption{Details of experimental equipment}
 \scalebox{0.8}{
  \begin{tabular}{|l|c|c|c|}
   \hline 
   Component&Manufacturer&Model number&Details\\ \hline \hline
   Semiconductor laser&NTT Electronics&NLK1C5GAAA&Wavelength of \SI{1547.79}{\nano\metre}, no isolator\\ \hline
   Photoreceiver&Thorlabs&RXM10AF&\SI{10}{\giga\hertz} bandwidth\\ \hline
   Oscilloscope&Tektronix&MSO72304DX&\SI{23}{\giga\hertz} bandwidth, \SI{50}{\giga Sample/\sec}\\ \hline
   Voltage-controlled variable attenuator&Thorlabs&V1550PA&Wavelength of \SI{1550}{\nano\metre}\\ \hline
   Optical circulators&Thorlab&CIR1550PM-APC&Wavelength of \SI{1520}{}-\SI{1580}{\nano\metre}\\\hline
   \multirow{2}{*}{Fiber couplers}&\multirow{2}{*}{Thorlabs}&PN1550R5A2&Wavelength of \SI{1550}{}$\pm$\SI{15}{\nano\metre}, 90:10 Split\\\cline{3-4}
   &&PN1550R5A1&Wavelength of \SI{1550}{}$\pm$\SI{15}{\nano\metre}, 50:50 Split\\ \hline
   Optical isolator&Thorlab&IOT-G-1550A&Wavelength of \SI{1550}{\nano\metre}\\
   \hline 
  \end{tabular}
  }
 \label{table2}
\end{table}

\begin{table}[ht]
 \centering
 \caption{The threshold injection current of each laser}
  \begin{tabular}{|c|c|c|}
   \hline 
   Laser&Threshold injection current value& Injection current value\\\hline\hline
   Laser 1A&\SI{11.0}{\milli\ampere}&\SI{12.0}{\milli\ampere}\\\hline
   Laser 1B&\SI{11.6}{\milli\ampere}&\SI{12.7}{\milli\ampere}\\\hline
   Laser 2A&\SI{11.7}{\milli\ampere}&\SI{12.9}{\milli\ampere}\\\hline
   Laser 2B&\SI{11.8}{\milli\ampere}&\SI{13.0}{\milli\ampere}\\\hline 
  \end{tabular}
 \label{table3}
\end{table}

%\bibliography{export}

\section*{Acknowledgements}
The authors thank A. Roehm for discussions about laser networks. This research was funded in part by the Japan Science and Technology Agency through the Core Research for Evolutionary Science and Technology (CREST) Project (JPMJCR17N2), and in part by the Japan Society for the Promotion of Science through Grant-in-Aid for Research Activity Start-up (22K21269), and Grant-in-Aid for Early-Career Scientists (23K16961),
Grants-in-Aid for Scientific Research (A) (JP20H00233) and Transformative Research Areas (A) (JP22H05197).

\section*{Author contributions statement}
T.M. conceived the numerical simulation and the experiment, H.I. conducted the numerical simulation and the experiment, H.I., T.M., and M.N. analyzed the results. All authors discuss the results, H.I. wrote the manuscript, and all authors reviewed the manuscript.

\section*{Additional information}
%\subsection*{Accession codes}
\subsection*{Competing interests}
The authors declare no competing interests.

\end{document}